# Indication of Meissner Effect in Sulfur-Substituted Strontium Ruthenates

A.M. Gulian[1,2*] and V.R. Nikoghosyan[2,3]

[1]*Physics Art Frontiers, Ashton, MD 20861*
[2]*Chapman University, Advanced Physics Laboratory, Burtonsville, MD 20866*
[3]*Institute for Physics Research of NAS, Ashtarak-2, 0203, Armenia*

*Abstract* – Ceramic samples of $Sr_2RuO_{4-y}S_y$ (y=0.03-1.2) with intended isovalent substitution of oxygen by sulfur have been synthesized and explored in the temperature range 4-300K. It is found that at a range of optimum sulfur substitution the magnetic response of ceramic samples reveals large diamagnetic signal with amplitudes approaching comparability with that of the YBCO-superconductors. Contrary to a pure ceramic $Sr_2RuO_4$, if properly optimized, the resistivity of sulfur-substituted samples has a metallic behavior except at lower temperatures where an upturn occurs. Both synthesis conditions and results of measurements are reported. The Meissner effect may point to high-temperature superconductivity.

## 1. Results

By the reasons which we will discuss later (in Section 3) we undertook synthesis of ceramic pellets of the composition $Sr_2RuO_{4-y}S_y$, with y = 0.03 – 1. For some range of y-values, y~0.3, the samples demonstrate very strong diamagnetic signals, about 10% of that of YBCO superconductors (Fig. 1).

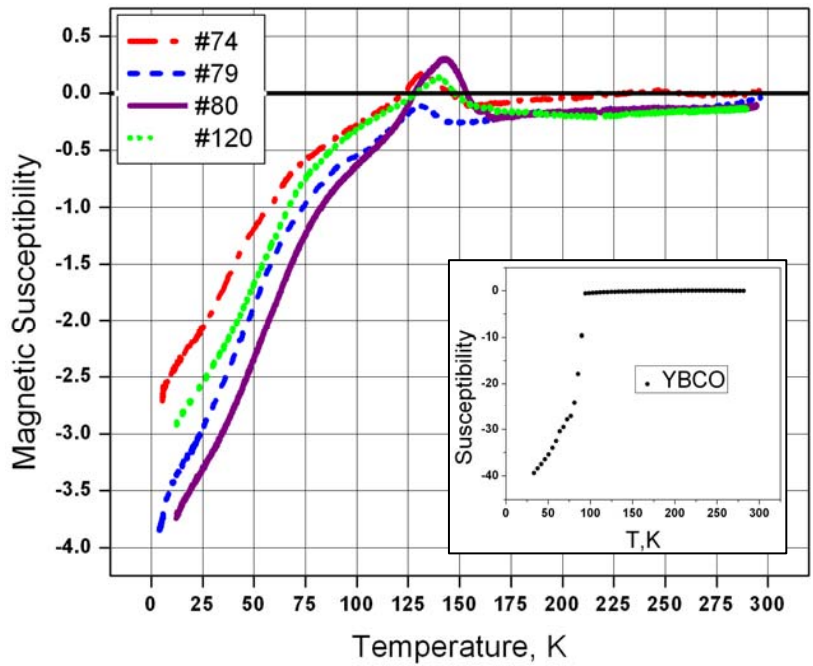

FIG.1. Magnetic susceptibility of samples with y=0.1 (#74 and #120), y=0.3 (#80), and y=0.4 (#79). Inset shows the result for the same size YBCO pellet measured by the same apparatus (Maxwell-Wien type double-coil bridge). No dc magnetic field is applied. The Earth's field was not screened.

*Corresponding author. Email: gulian@chapman.edu or gulian@hotmail.com

Submitted 9/25 2011

This magnetic response is quite stable against thermocycling, and has high reproducibility. In the next section we describe all essential sample preparation techniques.

## 2. Sample preparation

We undertook synthesis of the composition $Sr_2RuO_{4-y}S_y$, with y=0.03, 0.1, 0.3, 0.45, 0.65, 0.85 and 1.0 via traditional solid-phase synthesis route with initial ingredients $SrCO_3$ (99.994%), $RuO_2$ (99.95% purity), and SrS (99.995%); sometimes, with almost equal success, $SrSO_4$ (99.995%) was substituted for SrS. These components have been mixed in the (2-y):1:y molar weight proportion in the agate mortar (mixing duration about 1 hour per gram of weight), and annealed about 10 hours in air at the temperature $1000^0C$ using a muffle furnace (MTI SMF-1100). The temperature was changed at the speed about $4^0$/min. After this heat treatment the mixtures underwent grinding in the same mortar and subsequently pelletized (pellets ¼ inch in diameter, about 500 mg in weight). These pellets have been annealed in the same furnace, for 10 hours at $1000^0C$. The furnace heating took place at the speed $4^0$/min, and cooling at $100^0$/hour. During $1000^0C$ treatment the temperature was raised fast to $1150^0C$ (in 30 min), and lowered with the same speed three times. Our belief was that sulfur is partially replacing oxygen, and that this substitution is isovalent. Correspondence between this suggestion and the experiment was possible to test by careful weight control after each preparation procedure. Sometimes, though not always, it was almost ideal.[1]

    Ceramics of $Sr_2RuO_4$ have resistivity of semiconductor type in the whole scale of room to absolute zero temperatures (see, e.g., [2]), although high-quality single crystals of the same material have metallic behavior in the *(ab)*-plane and (below some temperature) along the *c*-axis [3]. Most likely, semiconductor intergranular links are responsible for this difference. We found that the addition of sulfur gradually facilitates the metallization. Moreover, the conductivity at room temperatures monotonically increases then drops as sulfur concentration increases. The concentration of sulfur that maximizes the conductivity at room temperature is designated "optimal". For sulfur concentrations close to the optimal, the dependence of resistance on temperature has metallic shape from the room to about 50-100K, and then has an upturn. For the best samples tested by us, these upturn values do not exceed the room-temperature level even at temperatures as low as 4K. The curves corresponding to different values of the parameter *'y'* are shown in Fig. 2.

---

[1] Perhaps, one should have used the terminology "isovalent substitution" more delicately. However, proper measurement of weights during different stages of sample preparation is assuring that this suggestion is not far from reality. For example, one of the samples (#121) having a composition $Sr_2RuO_{3.9}S_{0.1}$ had initial weight of ingredients 507mg (sulfur was introduced via $SrSO_4$), and final weight 402 mg, which corresponds to 20.7% of weight loss vs. 20.8% of theoretical weight loss at the isovalent substitution. In many cases this accordance was not as perfect. A 10% difference was typical. It is not excluded, however, that deviations from the theoretical estimates are playing a very important role in defining the physical properties of samples.

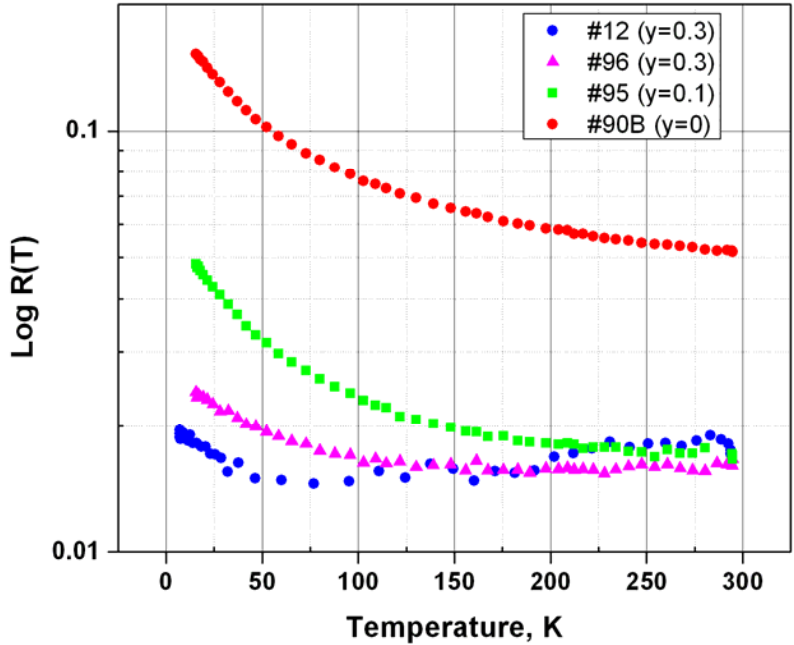

FIG. 2. Resistivity reduction and "metallization" by optizing values of '$y$'. This process is very sensitive to preparation conditions: minor differences strongly affect the conductivity, as clear from the comparison between samples #12 and #96. On the vertical scale, the units are such that R(300K) corresponds to the resistivity ~30 mOhm-cm for sample #90B.

### 3. Discussion

The results obtained here are connected with those reported in [4,5] (see also more complete publication [6]). An important factor, which was not known at the time of these earlier publications, is the presence of some amount of sulfur in the sample #L13-C128-G1 (see Refs. [5,6]) having simultaneous resistive and magnetic transitions with the onset as high as 250K. After many trials to reproduce this sample, we subjected it to Auger-analysis, which revealed the presence of sulfur in it. How sulfur ended up in that sample is a big puzzle, but its resolution is not the topic of this report. Rather, it is mentioned to explain our choice of the present composition.

With the introduction of sulfur, novel results began to appear rather quickly. Moreover, the reproducibility of the samples is relatively good, especially for the diamagnetic susceptibility, which does not depend essentially on intergranular linking. Reproducibility of the resistive curve is more delicate. Among other things, one should be aware of hygroscopic nature of $RuO_2$. Even powders containerized under argon may have up to 12% of moisture. For our initial series of samples we regretfully have not subjected $RuO_2$ to any thermal pre-annealing, but later found it to be very important for reproducibility (we recommend annealing of $RuO_2$ at $800^0$C for 3 hours, and then storing it in a vacuum desiccator: with this precaution no noticeable weight changes occur during some months of storing).

There is no need to say that our results are of a preliminary nature. To be highly cautious, one may expect that the Meissner effect, even with accompanying resistive transition, cannot conclusively establish the occurrence of superconductivity. The Ruddlesden-Popper family of materials: $Sr_{n+1}Ru_nO_{3n+1}$ (n=1,2,3,…), which is basic for our modified composition[2], may reveal very sophisticated metamagnetic properties, as well as resistive transitions (see, e.g., [8] ). However, to the best of our knowledge, these materials never demonstrated (large) diamagnetism. In principle, some orthovanadates have done that [9], but in response to applied kOe-scale fields which is not our case.

It would be interesting to continue and modify our approach, to reveal possibly hidden resistive transitions in accordance to a model excellently described lately by Kresin and Wolf [1]. The practical task is actually in breaking high-resistance (intergranular) barriers and allowing the proximity effect to reveal possible superconductivity. Also, one can try other substitutions instead of sulfur. *Se* and *As* are the most interesting candidates inspired by the wave of research with the pnictides.

**Acknowledgements**

We thank Dennis Winegarner for partial financial support of this work. We are grateful to many of our colleagues for useful discussions and encouragement. There are too many to list on this report.

---

[2] Interestingly, this Ruddlesden-Popper series has all the attributes of the substance considered promising for higher-temperature superconductivity from the recent viewpoint of theoretical analysis [7].